\documentclass[aps,pra,10pt,twocolumn,superscriptaddress,amsmath,amssymb,showpacs]{revtex4-1}

\usepackage{color}
\usepackage{xcolor}
\usepackage{soul}
\usepackage{tikz}
\usepackage{bm}
\usepackage{nicefrac}
\usepackage{amssymb}
\usepackage{graphicx}
\usepackage{bm}
\usepackage{braket}
\usepackage{mathrsfs}
\usepackage[normalem]{ulem}

\newcommand{\B}[1]{\mathbf{#1}}

\tikzstyle{mybox} = [draw=black, fill=grey, very thick,rectangle, rounded corners, inner sep=10pt, inner ysep=20pt]
\tikzstyle{fancytitle} =[fill=black, text=white]
\usetikzlibrary{arrows,shapes,trees, shadows}
\usetikzlibrary{decorations.pathmorphing}
\usetikzlibrary{decorations.markings}
\usetikzlibrary{shapes.geometric}
\usetikzlibrary{positioning}
\tikzset{partial ellipse/.style args={#1:#2:#3}{insert path={+ (#1:#3) arc (#1:#2:#3)}}}

\definecolor{UniBlue}{rgb}{0.0,0.29,0.6}
\definecolor{UniRed}{rgb}{0.757,0,0.165}
\definecolor{UniGray}{rgb}{.604,.608,.612}
\definecolor{LighterBlue}{HTML}{6E92B8}
\definecolor{LighterRed}{HTML}{E8582C}

\begin{document}

\title{Spectroscopic signatures of quantum friction}

\author{Juliane Klatt}
\affiliation{Physikalisches Institut, Albert-Ludwigs-Universit\"{a}t Freiburg, Hermann-Herder-Str. 4, D-79104, Freiburg i. Br., Germany}
\author{Robert Bennett}
\affiliation{Physikalisches Institut, Albert-Ludwigs-Universit\"{a}t Freiburg, Hermann-Herder-Str. 4, D-79104, Freiburg i. Br., Germany}
\author{Stefan Yoshi Buhmann}
\affiliation{Physikalisches Institut, Albert-Ludwigs-Universit\"{a}t Freiburg, Hermann-Herder-Str. 4, D-79104, Freiburg i. Br., Germany}
\affiliation{Freiburg Institute for Advanced Studies, Albert-Ludwigs-Universit\"{a}t Freiburg, Albertstr. 19, D-79104 Freiburg i. Br., Germany}

\date{\today}

\begin{abstract} 
We present a formula for the spectroscopically-accessible level shifts and decay rates of an atom moving at an arbitrary angle relative to a surface. Our Markov formulation leads to an intuitive analytic description whereby the shifts and rates are obtained from the coefficients of the Heisenberg equation of motion for the atomic flip operators but with complex Doppler-shifted (velocity-dependent) transition frequencies. Our results conclusively demonstrate that for the limiting case of parallel motion the shifts and rates are quadratic or higher in the atomic velocity. We show that a stronger, linear velocity-dependence is exhibited by the rates and shifts for perpendicular motion, thus opening the prospect of experimentally probing the Markovian approach to the phenomenon of quantum friction. \end{abstract}

\maketitle

\section{Introduction}

How does an atom with a fluctuating dipole moment behave when moving relative to a surface? Given the recent resurgence of interest in short-range fluctuation-induced forces brought about by advances in micro and nano-scale technology, one would expect this question to
have a clear-cut, unambiguous answer. Indeed the intuition for the effect is clear --- the properties of a fluctuating atomic dipole depend on the distance to an image dipole \cite{Casimir1948a}, meaning that a relative motion between the two should cause velocity-dependent
dynamical corrections. However, even for relatively simple and idealised models of atoms and surfaces there are significant disagreements between different approaches to calculating, for instance, the frictional force that an atom may experience while moving parallel to a surface. For example, Refs~\cite{Intravaia2014, Hoye2015} disagree with Refs~\cite{Persson1995, Schaich1981, Scheel2009b} about the power law governing the velocity dependence of the effect at zero temperature -- it is even argued in \cite{Philbin2009} that the effect does not exist at all, or in \cite{Intravaia2015} that some methods (e.g., \cite{Barton2010}) are very sensitive to the initial velocity preparation. These discrepancies arise largely because several different and incompatible formalisms have been used in calculating the velocity-dependent force. These include linear-response theory \cite{Dedkov2002}, Born-Markov approximations \cite{Scheel2009b}, time-dependent perturbation theory \cite{Intravaia2015} and appeals to a generalised fluctuation-dissipation theorem \cite{Intravaia2014}. As in all physics, the only real validation of a successful approach is via experiments, which are sorely lacking in atomic friction. This is because the forces involved are extremely small, and there are serious experimental challenges concerning precision measurements of forces on atoms near surfaces \cite{Sorrentino2009, Tarallo2012}, meaning that it is difficult to confirm or exclude particular theoretical approaches.

\begin{figure}
  \includegraphics{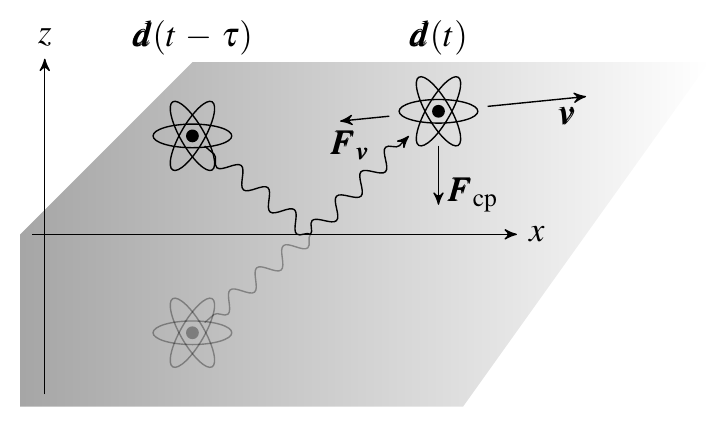}
  \caption{Atom moving next to a surface with velocity $\B{v}$. Its electric dipole $\B{d}(t)$ fluctuates about zero. The atom may have emitted a photon at time $t-\tau$ which is reabsorbed at time $t$.}
  \label{fig:setup}
\end{figure}

Here, we take a different route and consider the much more experimentally accessible internal dynamics of the atom, which in principle can be measured spectroscopically, thus providing a testable prediction of a velocity-dependent quantum-vacuum effect. We will present new results for the paradigmatic setup of a zero-temperature neutral atom with dipole moment $\B{d}$ and non-relativistic velocity $\B{v}$ moving next to a perfectly smooth macroscopic surface, as shown in Fig.~\ref{fig:setup}. 
For an atom at rest, the interaction of the atom's fluctuating dipole moment with its image causes the Casimir-Polder corrections to the atom's levels and decay rates \cite{Casimir1948a}. If the atom is allowed to move relative to the surface, fields induced by images at \emph{previous} times reach the atom -- in other words the motion of the atom causes it to `see' its image as being at a different position to where it is currently, resulting in dynamical effects.
 
\section{Model}
 
The dynamics shown in Fig.~\ref{fig:setup} consists of three mutually coupled parts: (i) the atom's center-of-mass motion, (ii) the internal dynamics of the atom and (iii) the dynamics of the medium-assisted electromagnetic field which surrounds the atom. The center-of-mass motion may be separated from the other degrees of freedom in the spirit of the Born-Oppenheimer approximation. Accordingly, the coupled atom-field dynamics are solved for a fixed atomic velocity. For the description of the dynamics of the composite field-matter system consisting of the electromagnetic field coupled to the charges making up the medium, we use the framework of macroscopic quantum electrodynamics \cite{Gruner1996a, Scheel2009}. The latter is a prescription for the quantisation of the electromagnetic field interacting with macroscopic, dispersive and absorbing bodies. As a consequence, the field-matter system is represented by a bosonic field with elementary excitations $\B{f}_\lambda$ for each electric or magnetic-type excitation $\lambda=\{e,m\}$, with position $\B{r}$ and frequency $\omega$. The Hamiltonian $H_\text{F}$ describing this part of the dynamics is then simply the canonical form integrated over all space;
\begin{equation}
H_\text{F} = \hbar  \int\!\! d^3\B{r} \int_0^\infty\!\!\!\!\!\!\! d\omega\omega\,\B{f}^\dagger_\lambda (\B{r},\omega)\cdot\B{f}_\lambda (\B{r},\omega)\,. 
\end{equation}
The free atom of mass $m$ and center-of-mass momentum $\B{p}$ is described by a Hamiltonian
\begin{equation}
H_\text{A}= \frac{\B{p}^2}{2m} + \sum_n E_n \ket{n} \bra{n}\,,
\end{equation}
where $n$ indexes an atomic level of energy $E_n$. The third and final part of the Hamiltonian comprises the interaction between the macroscopic QED electric field $\B{E}(\B{r})$  and the atom. This interaction is described in the dipole approximation by a Hamiltonian
\begin{equation}\label{eq:Hint}
H_{\text{AF}}=-\sum_{mn} \ket{m}\bra{n} \B{d}_{mn} \cdot \B{E}(\B{r}_\text{A})\,,
\end{equation}
furnishing us with the total Hamiltonian $H=H_\text{F}+H_\text{A}+H_{\text{AF}}$. Note that magnetic contributions to the interaction are omitted since they play a minor role in close proximity to the surface. The macroscopic QED electric field in a region with permittivity $\varepsilon(\B{r},\omega)$ and permeability $\mu(\B{r},\omega)$ is given explicitly in terms of the bosonic operators $\B{f}_\lambda(\B{r},\omega)$ introduced above by,
\begin{equation}\label{mqedEField}
\B{E}(\B{r})=\sum_\lambda \int\!\!d^3 \B{r}'\int_0^\infty\!\!\!\!\!\!\! d\omega\,\B{G}_\lambda(\B{r},\B{r}',\omega)\cdot \B{f}_\lambda (\B{r}',\omega)+\text{h.c.}
\end{equation}
with
\begin{align}
&\B{G}_\text{e}(\B{r},\B{r}',\omega)=i\frac{\omega^2}{c^2}\sqrt{\frac{\hbar\varepsilon_0}{\pi}\text{Im}\varepsilon(\B{r}',\omega)}\B{G}(\B{r},\B{r}',\omega)\,,\\
&\B{G}_\text{m}(\B{r},\B{r}',\omega)=i\frac{\omega}{c}\sqrt{\frac{\hbar}{\pi\mu_0}\frac{\text{Im}\mu(\B{r}',\omega)}{|\mu(\B{r}',\omega)|^2}}\B{G}(\B{r},\B{r}',\omega)\,,
\end{align}
where $\B{G}_\lambda(\B{r},\B{r}',\omega)$ is the Green's function for the Helmholtz equation
\begin{align}
\left[\nabla\times\frac{1}{\mu(\B{r},\omega)}\nabla\times-\frac{\omega^2}{c^2}\varepsilon(\B{r},\omega)\right] \B{G}_\lambda(\B{r},\B{r}',\omega)=\delta(\B{r}-\B{r}')\,.
\end{align}
This Green's function describes the propagation of field-matter excitations of frequency $\omega$ from $\B{r}'$ to $\B{r}$ thereby encoding all the information about the environment, i.e., its geometry as well as its dispersive and absorptive properties.

Using Eq.~\eqref{mqedEField} in our Hamiltonian $H$, we have for the Heisenberg equations of motion,
\begin{equation}
\dot{A}_{mn}(t) = i\omega_{mn}+\frac{1}{i\hbar}\left[{A}_{mn}(t)\,,\,H_\text{AF}(t)\right]\,, 
\end{equation}
for the atomic flip operators $A_{mn} \equiv \ket{m} \bra{n}$ a differential equation which can be formally solved in a Dyson-like expansion in the square of the electric dipole moment $\B{d}$ of the atom. The dipole operator $d_{mn}A_{mn}$ induces an atomic transition from one electronic level to another, which will necessarily be accompanied by the emission/absorption of a body-assisted field excitation given the form of the atom-field coupling in Eq. (\ref{eq:Hint}). Hence, restricting to quadratic order in $\B{d}$ corresponds to considering (at most) two emission or absorption events, which -- if a surface is present -- means neglecting multiple reflections. Doing this, we find for the dynamics of the $\B{d}^2$ approximation ${A}^{(2)}_{mn}(t)$ to the atomic flip operator 
\begin{align} \label{HeisenbergEOM}
\dot{A}^{(2)}_{mn}&(t) = \dot{A}^{(0)}_{mn} (t)\notag \\
&-\frac{1}{i\hbar} \sum_{ij} \left[{A}^{(0)}_{mn}(t) \, , \, {A}^{(0)}_{ij}(t) \B{d}_{ij} \cdot \B{E}^{(1)}(\B{r}_\text{A},t)   \right]\, . 
\end{align}
where $\B{E}^{(1)}$ is the free field plus that induced by an atom described via the $\B{d}^0$ approximation ${A}^{(0)}_{mn}(t)$ to the atomic flip operator. Taking the normal-ordered vacuum expectation value of \eqref{HeisenbergEOM} and utilising the Heisenberg equation of motion for the $\B{f}_\lambda(\B{r},\omega)$ one arrives at
\begin{equation} \label{HeisenbergEOMVEV}
\langle \dot{A}^{(2)}_{mn}(t)\rangle  =\left\{ i\omega_{mn} -\left[C_n(t)+C^*_m(t) \right] \right\} \langle {A}^{(2)}_{mn}(t)\rangle \; ,\end{equation}
where we have replaced ${A}^{(0)}_{mn}(t) \to {A}^{(2)}_{mn}(t)$ on the right-hand side. The resulting error will be of order $\B{d}^4$, as can be easily seen from the coefficients $C_n= \sum_k C_{nk}$ given explicitly by \cite{Buhmann2004},
\begin{align} \label{Cnk}
C_{nk} = \frac{\mu_0}{\pi\hbar} \int_{t_0}^t\!\!\!\! dt' \int_0^\infty\!\!\!\!\!\!\! d\omega \omega^2 \B{d}_{nk} \cdot & \text{Im} \B{G}(\B{r}_\text{A},\B{r}_\text{A}',\omega) \cdot \B{d}_{kn} \notag \\
& \times e^{-i(\omega-\omega_{nk})(t-t')} \; , 
\end{align}
where $\B{r}_\text{A}=\B{r}_\text{A}(t)$ and $\B{r}_\text{A}'=\B{r}_\text{A}(t')$ are the current and previous position of the atom, respectively. Here we have used a well-known integral relation for electromagnetic dyadic Green's functions \cite{Buhmann2004}:
\begin{equation}
\sum_{\lambda} \int\!\! d^3 \B{s}\,\B{G}_{\lambda}(\B{r}, \B{s}, \omega)\cdot \B{G}^*_{\lambda}(\B{s}, \B{r}', \omega)=\frac{\hbar\mu_0}{\pi} \omega^2 \text{Im} \B{G}(\B{r},\B{r}',\omega)\,.
\end{equation}
Inspection of Eq.~\eqref{HeisenbergEOMVEV} shows that the real and imaginary parts of $C_n$ deliver respectively the rate of spontaneous decay $\Gamma_n$ and the level shift $\hbar\delta\omega_n$ with respect to the bare level $E_n$ of the state $n$ via
\begin{equation}\label{RatesShiftsReIm}
\Gamma_n = 2\sum_{k<n} \text{Re} C_{nk} \quad,\quad \delta\omega_n = \sum_k  \text{Im} C_{nk}\,.
\end{equation}
 
Having set up the model, we now present our main results, which are the first predictions of level shifts and decay rates for an atom moving in an arbitrary direction near a surface. In order to produce concrete numbers for the level shifts and decay rates, we employ a Markov approximation in which the coefficients in \eqref{HeisenbergEOMVEV} are presupposed to be time-independent. In other words, we assume clear separation of the three timescales involved. Firstly, the dynamics of both the field and the internal of the atom are assumed to happen in a much faster pace than the atomic center-of-mass motion. Hence, the atom's position and velocity may be treated as instanteneous and fixed -- eliminating implicit time dependences in the $C_{nk}$. Secondly, typical timescales of the field's dynamics -- given by its memory, i.e. auto-correlation time -- are presupposed to be very small compared to the timescales on which electronic transitions in the atom take place. Therefore, any residual time-dependence -- saturated on the scale of the field's memory -- will not be resolved in the internal atomic dynamics. This is the well-known coarse-graining effect the Markov approximation relies on. Consistency with such an approximation requires that we assume approximately uniform motion $\B{r\vphantom{r'}}_\text{A}-\B{r}'_\text{A} \approx \B{v}(t-t') \equiv \B{v}\tau$.

We take advantage of translational invariance parallel to the surface to take the Fourier transform $ \underline{\text{Im} \B{G}}$ of the imaginary part of the Green's function appearing in Eq.~\eqref{Cnk}. Similarly, we split up the atomic velocity $\B{v}$ and the wave vector $\B{k}$ into components parallel to the surface $\{\B{v}_\parallel,\B{k}_\parallel\}$ and perpendicular to it $\{\B{v}_\perp,\B{k}_\perp\}$, giving for Eq.~\eqref{Cnk},
\begin{align}\label{CnkParallelAndPerp}
C_{nk} =& \frac{\mu_0}{\pi\hbar} \int_0^\infty \!\!\!\!\!\!\!d\tau \!\!\int_0^\infty\!\!\!\!\!\!\! d\omega \omega^2\!\!\int\!\! d^2  \B{k}_\parallel  \notag \\
& \times \B{d}_{nk} \cdot  \underline{\text{Im} \B{G}}(\B{k}_\parallel,{z}_\text{A},\omega) \cdot \B{d}_{kn} e^{-i(\omega-\omega_{nk}')\tau}\, ,
\end{align}
where a Doppler shifted frequency $\omega_{nk}' \equiv \omega_{nk} + \B{k} \cdot \B{v}$ has naturally arisen and we have made use of a shorthand $\B{G}(\B{k}_\parallel, z,\omega)\equiv \B{G}(\B{k}_\parallel, z,z,\omega)$.  Finally, we have taken the limit $t_0\to -\infty$, which is justified as long as $t_0$ is significantly larger than the width of the field's memory kernel, consistent with the Markov approximation.

Since we ultimately want determine the shifts $\delta\omega_n$ and rates $\Gamma_n$ given in (\ref{RatesShiftsReIm}) and accordingly aim to identify the real and imaginary parts of (\ref{CnkParallelAndPerp}),  it is useful to further simplify $\underline{\text{Im}\B{G}}$ (the Fourier transform of the imaginary part of $\B{G}$) as this quantity has no obvious separation into real-valued and imaginary components . To this end, we note that for real $\B{d}_{nk}$ only the symmetric portion $\mathcal{S} \underline{\B{G}}$ of the Fourier-transformed Green's tensor $ \underline{\B{G}}$ contributes, which, for a half-space geometry described by $\B{G}=\B{G}^\text{HS}$ that we shall use later on, is precisely the part for which Fourier transforming and taking the imaginary part commute: $ \mathcal{S} \left[\underline{\text{Im}\B{G}^\text{HS}}\right]= \mathcal{S} \left[\text{Im}\underline{\B{G}^\text{HS}}\right]$. Now we have the imaginary part of the Fourier transform (rather than vice versa) which is manifestly real. Thus we now have a clear separation of real and imaginary parts in \eqref{CnkParallelAndPerp}, enabling us to easily identify level shifts and rates of spontaneous decay via Eq. (\ref{RatesShiftsReIm}). 

Furthermore we shall specialise to the non-retarded (i.e. near-field) regime where the atom-surface distance $z_\text{A}$ is short enough that the finite round-trip time of a reflected photon is negligible compared to atomic timescales. This regime is defined by $\omega_{nk} z_\text{A}/c \ll 1$. Under these conditions, $k_\perp \approx (-k_\parallel^2)^{1/2}$ and the Doppler shifted atomic transition frequencies become $\omega_{nk}'\to \omega_{nk} + \B{k}_\parallel \cdot \B{v}_\parallel-i k_\parallel v_{\perp}$, where we have made the physical choice of branch of the square root such that evanescent waves are decaying away from the surface as $z\!\to\!\infty$.

Lastly, let us spell out in more detail the connection between contributions stemming from a finite atomic center-of-mass velocity $\B{v}$ to the internal atomic dynamics on the one hand and quantum friction, i.e. finite-$\B{v}$ contributions to the Casimir-Polder force, on the other hand. An agreed-upon feature of the various approaches to quantum friction mentioned in the introduction is the following expression for the Casimir-Polder force:
\begin{align}\label{eq:force}
\B{F}_\text{CP}(t)&=\frac{i\mu_0}{4\pi^3}\,\nabla\!\!\int_0^t\!\!\!\!dt'\!\!\int\!\!d^2\B{k}_\parallel\!\int_0^\infty\!\!\!\!\!\!\! d\omega\omega^2e^{-i\omega(t-t')}\\\nonumber
&\times\text{tr}\left[\mathcal{C}_d(t,t';\B{v})\cdot\text{Im}\B{G}(\B{k}_\parallel,{z}_\text{A},\omega)\right]e^{i\B{k}_\parallel\cdot(\B{r}_\text{A}-\B{r}'_\text{A})},
\end{align}
where $\mathcal{C}_d$ is the two-time correlator of the atomic dipole moment,
\begin{equation}
\mathcal{C}_d(t,t';\B{v})=\langle\B{d}(t)\B{d}(t')\rangle
\end{equation}
The Casimir-Polder force (\ref{eq:force}) experienced by an atom which moves parallel to a macroscopic surface, comprises the aforementioned dynamical contributions in two-fold manner. Firstly, \textit{explicitly} via the distance $\B{r}_\text{A}-\B{r}'_\text{A}$ travelled by the atom during emission at time $t$ and reabsorption at time $t'$ of a photon and secondly, \textit{implictly}, via the time evolution of the dipole operator which is evolves according to the entire Hamiltonian which naturally includes the atomic center-of-mass motion and hence $\B{v}$. This implicit dependence is indicated by the third argument of the correlator $\mathcal{C}_d$ and corresponds exactly to the finite-velocity contributions to the internal dynamics provided by Equations (\ref{RatesShiftsReIm}) and (\ref{CnkParallelAndPerp}).

There is consensus that the leading-order in $\B{v}$ contributions to the friction force acting on an atom moving parallel to the surface stem from the \textit{explicit} velocity dependence rather than the \textit{implicit} one in the correlator. Non-compatible assumptions on the precise long-time behaviour of the latter is nevertheless believed to bring about the contradicting results for that very leading order in relative velocity of the friction force. While Intravaia et~al. for instance assume a power-law decay of correlations for very large times \cite{Intravaia2014}, the Markovian approach presupposes exponential decay of correlations on all timescales \cite{Scheel2009b}. This large-time behaviour strongly influences the low-frequency contributions to quantum friction, which are the ones most sensitive to the \textit{explicit}, Doppler-shift like, corrections in Eq. (\ref{eq:force}). 

While \textit{not} lending our voice to either of the contradicting assumptions, we solely focus on the fact that the Markov approach -- in contrast to the generalised fluctuation-dissipation approach --  does not only render a prediction for dynamical corrections to the static Casimir-Polder force, but moreover predicts dynamical corrections on the level of the internal dynamics of the atom, associated with the inplicit velocity dependence of that force. The latter can be probed spectroscopically -- which, though challenging, is less demanding than a force measurement. Hence, the question whether the Markov approximation is legitimate for a Casimir-Polder setup subject to relative motion may in principle be answered by means of spectroscopy. The remainder of this work focuses on exactly that venture.

\section{Results}

In order to arrive at physical predictions, we now make use of the explicit non-retarded half-space Green's function (see, for example \cite{tai1994dyadic}) 
\begin{align}\label{NonRetGF}
\B{G}^\text{HS}(\B{r},\B{r}',\omega) = \frac{r_p(\omega) c^2}{8\pi^2 \omega^2}& \int_0^{2\pi} d\phi \int_0^\infty d\kappa \kappa^2 \notag \\
&\!\!\!\!\!\times e^{i\B{k_\parallel}\cdot(\B{r}_\parallel - \B{r}'_\parallel)}e^{-\kappa (z+z')} \B{a}\otimes \B{a}\,,
\end{align}
where $\B{a} = (\cos \phi, \sin \phi, i)$ and $r_p(\omega)=\frac{\varepsilon(\omega)-1}{\varepsilon(\omega)+1}$ is the non-retarded limit of the Fresnel reflection coefficient for $p$-polarized (transverse magnetic) radiation of frequency $\omega$ incident upon a non-magnetic $[\mu(\B{r},\omega)=1]$ half-space of permittivity $\varepsilon(\omega)$. We have written the frequency integral in Eq.~\eqref{NonRetGF} polar co-ordinates $\B{k}_\parallel = ( \kappa \cos \phi, \kappa\sin \phi )$. Defining a weighted squared dipole moment $d^{2(\phi)}_{nk} \equiv  \mathbf{d}_{nk} \cdot \left[\mathbf{a} \otimes  \mathbf{a}\right] \cdot  \mathbf{d}_{kn}$ and inserting \eqref{NonRetGF} into \eqref{CnkParallelAndPerp} with $\B{G}=\B{G}^\text{HS}$ and making use of the the Heaviside step function $\Theta(x)$ we find
\begin{align}\label{CnkMainEq}
C_{nk} = -\frac{i}{8\pi^2 \epsilon_0 \hbar} &\int_0^\infty \!\!\!d\kappa  \kappa^2  \int_0^{2\pi}\!\!\! d\phi \bigg[ r_p(\omega_{nk}')\Theta\left[\text{Re}(\omega_{nk}')\right] \notag \\
-\frac{1}{\pi}& \int_0^\infty d\xi \frac{\omega_{nk}'r_p(i\xi)}{\xi^2+\omega_{nk}'^2}\bigg]e^{-2\kappa z} d^{2(\phi)}_{nk} \, ,
\end{align}
which is our main result.  Its detailed derivation (see Appendix) proves that Eq.~\eqref{CnkMainEq} is valid for either sign of $v_\perp$, as long as the component of velocity away from the interface is not too large, as then the atom would `remember' having emerged from inside the medium, where our model does not apply. We also note that, in practice, the argument $\text{Re}[\omega_{nk}']$ of the step function in Eq.~\eqref{CnkMainEq} is dominated by $\omega_{nk}$, because $\omega_{nk} \gg {k_\parallel v_\parallel}$. To see this we note that the $k_\parallel$ integral in Eq.~\eqref{CnkMainEq} is effectively cut off at $\sim 1/z$. Then one can  easily check that the resulting condition $\omega_{nk} \gg { v_\parallel/z}$ is comfortably satisfied for all non-relativistic velocities and distances greater than a few nm. Equation \eqref{CnkMainEq} contains a remarkable amount of information --- the decay rates and frequency shifts for an atom with any velocity vector $\B{v}$ can be obtained from it simply by taking real and imaginary parts via Eqs.~\eqref{RatesShiftsReIm}. 

Physical insight can be gained from expanding our formula \eqref{CnkMainEq} in a Taylor series for low atomic velocities;
\begin{align}
 &\!\!\!\!C_{nk}^{\parallel \text{res} }\!\simeq\!\frac{-i}{ 32 \pi \epsilon_0 \hbar z^3  }\Bigg[ d^{{(i)}2}_{nk} r_p(\omega_{nk}) +\frac{3 d^{{(a)}2}_{nk}v_\parallel^2 }{8 z^2}r_p''(\omega_{nk})\Bigg]\!,  \label{CnkSmallVPara}\\
&\!\!\!\!C_{nk}^{\perp \text{res}} \!\simeq\! \frac{-id^{{(i)}2}_{nk}}{ 32 \pi \epsilon_0  \hbar z^3}\!\!\left[ r_p(\omega_{nk})\!-\!\frac{3i v_{\!\perp} }{2 z}r_p'(\omega_{nk})\right]\!\!, \label{CnkSmallVPerp}
\end{align}
where $d^{{(i)}2}_{nk} = d_{nk,x}^2+d_{nk,y}^2+2d_{nk,z}^2$ and  $d^{{(a)}2}_{nk} = 3d_{nk,x}^2+d_{nk,y}^2+4d_{nk,z}^2$  and the primes denote derivatives with respect to frequency. Here we have presented only the resonant part of the interaction since the non-resonant part is orders of magnitude smaller -- as shall be shown more explicitly later on. 

If applied to parallel motion and a plasma-model medium, Eq. (\ref{CnkSmallVPara}) exactly coincides with known results \cite{Ferrell1980}. We immediately see from Eq.~\eqref{CnkSmallVPara} [via Eq.~\eqref{RatesShiftsReIm}] that the lowest-order velocity-dependent corrections to the resonant level shifts  $\delta \omega_{n}^{\parallel}$ and decay rates $ \Gamma^{\parallel}_{n}$ for parallel motion are \emph{quadratic} in the atomic velocity --- in fact all odd-order terms vanish. This is expected given that the sign of the velocity should not matter for motion parallel to the surface, since the system is translationally invariant along those directions. Turning our attention to perpendicular motion, we observe from \eqref{CnkSmallVPerp} that the leading velocity-dependent corrections are linear in the velocity. This is physically reasonable as the system is not translationally invariant along the direction perpendicular to the surface, so that changing the sign of the velocity in that direction \emph{should} matter. Note, that the vanishing of all even orders in velocity in the case of parallel motion is by no means a contradiction to the fact that the friction force Eq.~(\ref{eq:force}) must be odd in relative velocity. As mentioned when this force was introduced, its leading-order in $\B{v}$ contribution does \textit{not} stem from the internal dynamics of the atom, i.e. the shifts and rates we studied in this section. Instead, leading order dynamical contributions to the friction force rather stem from an explicit, Doppler-shift like, $\B{v}$ dependence attributed to the distance $\B{r}_\text{A}-\B{r}'_\text{A}$ travelled by the atom during emission and reabsorption of a photon.

\section{Experimental Relevance}

\begin{figure}
\includegraphics[width = \columnwidth]{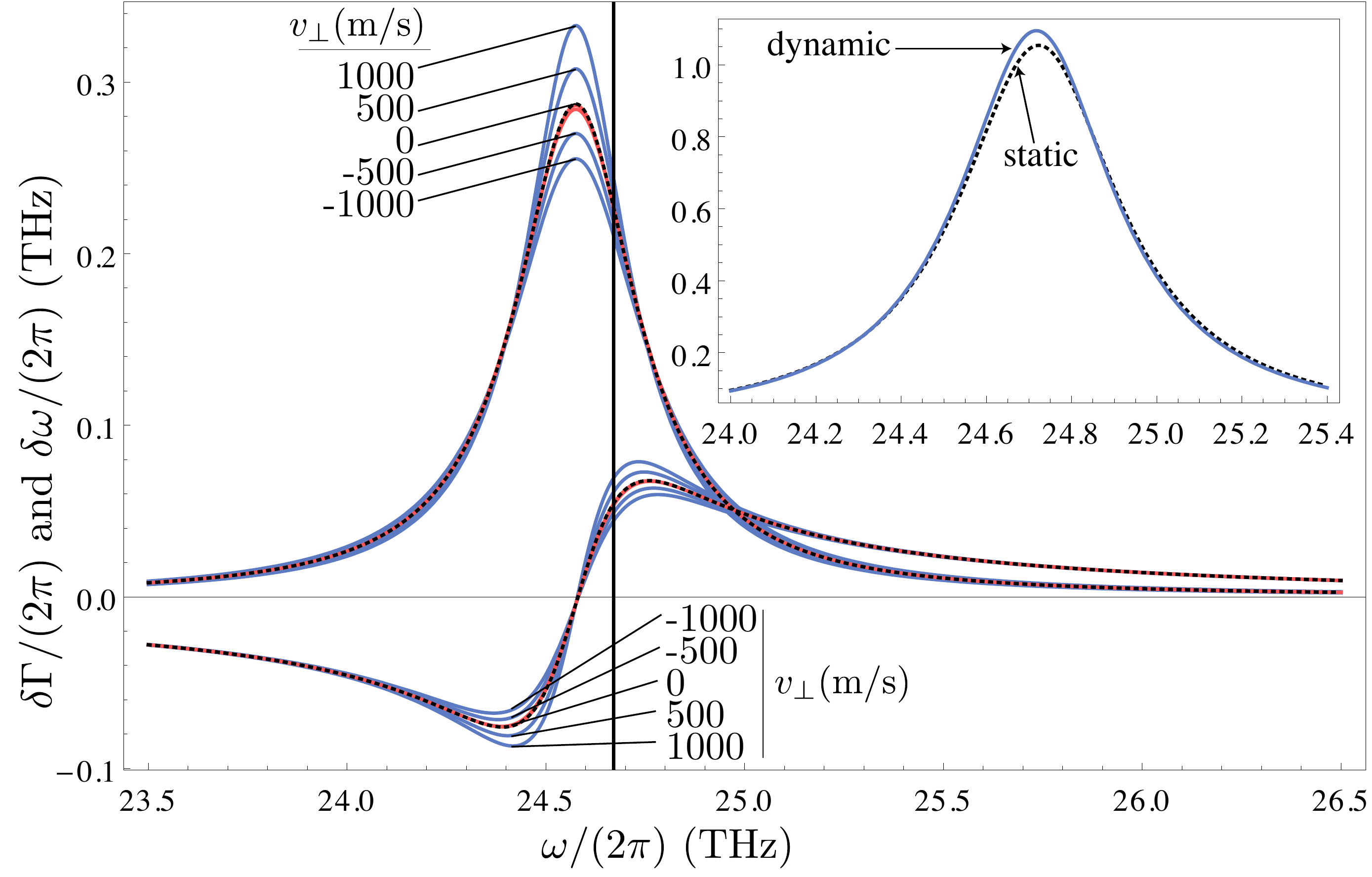}
\caption{Dependence of shifts $\delta\omega$ and decay rates $\Gamma$ (relative to the respective free-space values) on possible atomic transition frequencies of a $^{133}$Cs atom moving parallel (red) or perpendicular (blue) to a sapphire surface. The actual $6D_{3/2}\to7P_{1/2}$ transition is indicated by the vertical axis. The parameters describing the surface and the atom are $z_\text{A}\!\!=\!\!10$nm, $v\!\!=\!\!\pm500$m/s,$\pm1000$m/s, $\eta\!\!=\!\!2.71$, $\omega_\text{T}\!\!=\!\!1.56\!\cdot\!10^{14}$s$^{-1}$, $\omega_\text{P}\!\!=\!\!1.2\omega_\text{T}$, $\gamma\!\!=\!\!0.02\omega_\text{T}$ and $d\!\!=\!\!5.85\!\cdot\!10^{-29}$Cm and isotropic. We also include the static shifts and rates as dashed lines. The vertical line marks the actual transition frequency. The inset shows the  emission-line profile for static (dashed) and moving (blue) atoms at $v_{\!\perp}\!\!=\!\!500$m/s after averaging over $5\text{nm}\!<\!z_\text{A}\!<\!1\mu$m. For parallel motion the corrections are much smaller than for perpendicular one and have no visible effect on the line profile.}
\label{ParallelGraph}
\end{figure}

\begin{figure}
\includegraphics[width = \columnwidth]{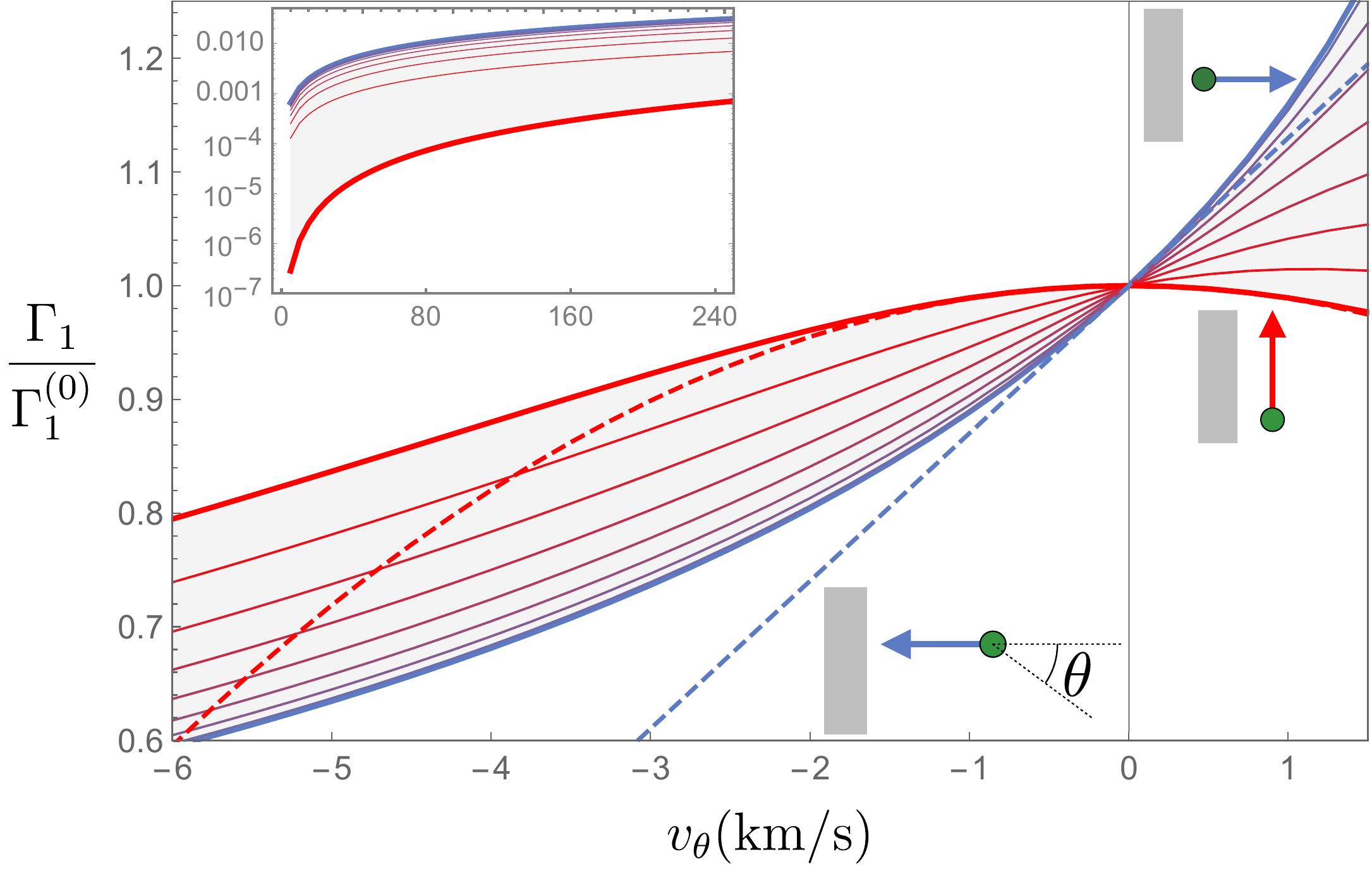}
\caption{Velocity-dependence of the decay rate for parallel (thick red line, $\theta\!=\!\pm\pi/2$) and perpendicular (thick blue line, $\theta\!=\!0,\pi$) motion of a $^{133}$Cs atom in front of sapphire in units of the static decay rate $\Gamma_1^{(0)}$, alongside the leading-order in $v$ expressions (dashed lines, with the parallel motion asymptote being barely distinguishable from the exact result at this scale). The thin lines between these extremal cases are for intermediate $0\!<\!\theta\!<\!\pi/2$ evenly spaced in steps of $\pi/16$. All parameters chosen here are the same as in  Fig.~\eqref{ParallelGraph}, and the transition frequency is taken as $\omega_\text{L}$ as indicated in Fig.~\eqref{ParallelGraph}. Inset: Detail of low-velocity region.}
\label{VelocityGraphShift}
\end{figure}

As a concrete example, consider $^{133}$Cs whose far infrared $6D_{3/2}\!\!\rightarrow\!\!7P_{1/2}$ transition is near-resonant with the $12.21\mu$m phononic resonance of ordinary sapphire \cite{Ducloy1995} which strongly enhances resonant Casimir-Polder effects. We describe the sapphire with a dominant-resonance Drude-Lorentz model, $\varepsilon(\omega) = \eta[1-\omega_\text{P}^2/(\omega^2-\omega_\text{T}^2+i\gamma\omega)]$, where $\omega_\text{P}$ is the plasma frequency, $\omega_\text{T}$ is an absorption line frequency, $\gamma$ is the damping parameter and $\eta$ accounts for the small background stemming from other atomic transitions. By means of (\ref{CnkMainEq}) we can now determine the velocity-dependent shifts and rates corresponding to this $^{133}$Cs transition in front of a sapphire surface. In Fig.~\ref{ParallelGraph} we plot the dependence of these shifts and rates on the atomic transition frequency for a selection of center-of-mass velocities. For parallel motion the dynamical corrections are much smaller than those for perpendicular motion. Hence, the inset in Fig.~\ref{ParallelGraph} depicts a spatially averaged ($5\text{nm}\!<\!z_\text{A}\!<\!1\mu$m) profile of the mentioned emission line -- as e.g. obtained by evanescent-wave spectroscopy -- of atoms moving \textit{perpendicularly towards} the surface at 500m/s. Compared to the static profile it is slightly shifted and clearly more peaked. An observation of the latter effect is demanding but much more in-reach than measurement of quantum friction forces. Similar experiments have already been carried out in order to measure the static Casimir-Polder shift \cite{Chevrollier1992}.
In Fig.~\ref{VelocityGraphShift} we show the velocity-dependence of the decay rate for a $^{133}$Cs atom moving arbitrarily with respect to the sapphire surface.

The non-resonant shifts and rates are a factor of $(\omega_\text{P}^2+\omega_\text{T}^2)^{1/2}/\gamma\!\!\approx\!\!100 $ smaller than the resonant shifts for any realistic choice of parameters, meaning that we can safely ignore them here. The known static, resonant, Casimir-Polder shifts and rates emerge from the terms zeroth order in velocity in Eqs.~\eqref{CnkSmallVPara} and \eqref{CnkSmallVPerp}. For an atom whose dipole moment is aligned along the positive $\B{z}$ direction they read
\begin{align} 
\delta \omega_{n,\Omega_+}^{(0)} &= \frac{ d_z^2  }{16\pi \epsilon_0    \hbar}\frac{\eta}{(\eta+1)}\frac{ \omega_p^2}{\omega_{\text{L}}\gamma }\frac{1}{z_\text{A}^3}\label{StaticShift}\,,\\
\Gamma_{n,\text{L}}^{(0)}&=\frac{ d_z^2  }{4\pi \epsilon_0    \hbar}\frac{\eta}{(\eta+1)}\frac{ \omega_p^2}{\omega_{\text{L}}\gamma }\frac{1}{z_\text{A}^3}\label{StaticRate}\,,
\end{align}
when evaluated at their respective maxima $\Omega_+$ and $\omega_\text{L}$ (see Tab.~\ref{table}) and taken at leading order in $\gamma \ll \omega_\text{T}$.

In Tab.~\ref{table} we summarise the lowest-order velocity dependent contributions to the level shifts and decay rates, expressed as ratios to the static quantities \eqref{StaticShift} and \eqref{StaticRate}.
\begin{table}
\begin{tabular}{c|c|c}
                                                              & Perpendicular motion                                          & Parallel motion \\\hline
$\delta\omega^\text{loc}_{n}/\delta\omega_{n,\Omega_+}^{(0)}$ & $\pm\frac{3v_\perp\vphantom{v_\parallel}}{\gamma z_\text{A}}$ & $-\frac{3v_\parallel^2}{\gamma ^2z_\text{A}^2}$ \\
$\Gamma^{\text{loc}}_{n}/\Gamma_{n,\text{L}}^{(0)}$           & $\frac{3v_\perp\vphantom{v_\parallel}}{\gamma z_\text{A}}$    & $-\frac{6v_\parallel^2}{\gamma ^2z_\text{A}^2}$ \\
    \end{tabular}
\caption{Leading-order contributions $\delta\omega^\text{loc}_{n}$ and $\Gamma^\text{loc}_{n}$ to resonant level shifts and rates for an atom moving with velocity $v\!\!<\!\!\gamma z_\text{A}$ next to a surface, with its dipole moment aligned along the positive $\B{z}$ direction. We have reported only the leading terms in $\gamma$ since the next terms will be smaller by a similar factor as the non-resonant parts, which we have ignored here. Each quantity is evaluated at the maximal points of the static quantities, that is, decay rates are evaluated at $\omega_\text{L} =\sqrt{\eta\omega_\text{P}^2/(\eta+1) +\omega_\text{T}^2}$ and level shifts at $\Omega_{+}=\omega_\text{L}+\gamma/2$.}
\label{table}
\end{table}
One can estimate the radius of convergence of the Taylor expansion by finding the ratios of sucessive orders. So from Tab.~\ref{table} one can see that the series converge for $v_{\parallel, \perp} \lesssim \gamma z_\text{A} \approx 1\text{THz}\cdot 1\text{nm} = 10^3$m/s. Typical velocities of atomic beams generated through thermal effusion are in the range $10^2-10^3$m/s (see, for example, \cite{Berman1997}), meaning that even the simple asymptotic formulae in Tab.~\ref{table} are immediately relevant to experiment.

Finally, let us discuss our assumptions and associated errors. Non-relativistic ($v/c\!\simeq\!10^{-5}$), non-retarded ($\omega_\text{T}z/c\!\simeq\!10^{-3}$), and single-reflection ($d^2\omega_\text{T}^2/\hbar\varepsilon_0c^3\!\simeq\!10^{-8}$) approximations lead to a relative error of about $10^{-3}$, which is not detectable for the class of experiments we compare to here. The Born-Oppenheimer and Markov approximations assume separation of field auto-correlation time, $\tau_\text{F}\!=\!\gamma^{-1}$, internal atomic time scales, $\tau_\text{A}\!=\!\Gamma^{-1}$, and center-of-mass time scales, $\tau_\text{C}$, respectively. The significant difference between the mass of the electron and the nucleus causes $\tau_\text{C}$ to clearly separate from internal atomic as well as field time scales ($\tau_\text{C}\gg\tau_\text{A},\tau_\text{F}$). However, the separation of the latter ($\tau_\text{A}\gg\tau_\text{F}$) strongly depends on $z_\text{A}$. The proposed experiment may hence serve to confirm or refute the applicability of the Markov approximation in this cross-over regime. Lastly, finite temperature enhances both static and dynamic effects by a factor $n(\omega_\text{A})+1$, where $n(\omega_\text{A})$ is the thermal occupation number of the mode $\omega_\text{A}$ corresponding to the atomic transition of interest. For the aforementioned transition of $^{133}$Cs at room temperature, $n(\omega_\text{A})\lesssim0.02$.

\section{Summary}

Here we have presented new, spectroscopically-accessible analytical predictions of the dynamical corrections to the internal structure of an atom as it moves in an arbitrary direction near a surface. We have obtained the general formula \eqref{CnkMainEq} that gives the full set of level shifts and decay rates for an obliquely moving, possibly excited, atom near a half-space with results shown in Fig.~\ref{VelocityGraphShift}. Our asymptotic results show that the relevant expansion parameter for small velocities is $v/(\gamma z_\text{A})$, which is large compared to, for example, $v/c$ or $v/(\omega_\text{T}z_\text{A})$. This, alongside the fact that the new results we have presented for perpendicular motion are linear in this parameter (in contrast to the quadratic dependence for parallel motion), means that these quantities are larger than previously thought, and therefore more easily measurable. In addition to being a velocity-dependent vacuum effect in its own right, our results constitute a testable prediction related to the less-accessible phenomenon of quantum friction. Our results represent a novel testbed for the applicability of the Markov approximation in this setting. Refuting Markovianity by experiments would rule out one of the contradicting standpoints in the quantum friction debate.

We would like to thank D.A.R.~Dalvit, M.B.~Far\'{i}as, S.~Scheel and B.~von~Issendorff for discussions. This work was supported by the DFG (Grants BU 1803/3-1476, GRK 2079/1) and the Freiburg Institute for Advanced Studies.

\appendix

\section{Derivation of Equation \eqref{CnkMainEq}}

Starting with the Heisenberg coefficients (6), substituting the non-retarded scattering Green's function for a half-space (7), and performing the $\kappa$ integration, one arrives at
\begin{align}\nonumber
C_{nk}=&\frac{1}{4\pi^3\hbar\varepsilon_0}\int\limits_0^T\!\!d\tau\!\!\int\limits_0^\infty\!\!d\omega\!\!\int\limits_0^{2\pi}\!\!d\phi\,\bm{d}_{nk}^{(\phi)2}\,\text{Im}r_\text{p}(\omega)e^{-i(\omega-\omega_{nk})\tau}\\\label{nonretarded}
&\times\left(2z_\text{A}-v\tau\cos\theta-i v\tau\sin\theta\cos\phi\right)^{-3}\,.
 \end{align}
Here, without loss of generality, the coordinate system is chosen such that the $y$-component of the atom's velocity is zero. Expanding the denominator of (13) in the unitless parameter $s=v\tau/2z_\text{A}$ around zero and abbreviating $f_{\phi,\theta}=\cos\theta+i\sin\theta\cos\phi$ yields
\begin{align}\nonumber
C_{nk}=&\frac{1}{64\pi^3\hbar\varepsilon_0z_\text{A}^3}\int\limits_0^T\!\!d\tau\!\!\int\limits_0^\infty\!\!d\omega\!\!\int\limits_0^{2\pi}\!\!d\phi\,\bm{d}_{nk}^{(\phi)2}\text{Im}r_\text{p}(\omega)e^{-i(\omega-\omega_{nk})\tau}\\
&\times\sum_{j=0}^\infty\frac{(j+2)!}{j!}s^jf^j_{\phi,\theta}\,.
\end{align}
Due to the oscillating nature of the integrand, the latter does not contribute to the integral for $\tau\!\gg\!\omega_{nk}$. Hence, the domain where the above series is convergent -- i.e., for $v\omega_{nk}\!\ll\!2z_\text{A}$ -- matches the domain where the integrand contributes. The powers of $\tau$ can be rewritten as derivatives with respect to $\omega$ which, via partial integration, may be shifted onto the reflection coefficient $r_\text{p}(\omega)$. Afterwards, the $\tau$-integral can be solved, giving,
\begin{align}\nonumber
C_{nk}&=\frac{1}{64\pi^3\hbar\varepsilon_0z_\text{A}^3}\sum_{j=0}^\infty\frac{(j+2)!}{j!}\int\limits_0^\infty\!\!d\omega\!\!\int\limits_0^{2\pi}\!\!d\phi\,\bm{d}_{nk}^{(\phi)2}\text{Im}r^{(j)}_\text{p}(\omega)\\
&\times\left(-\frac{ivf_{\phi,\theta}}{2z_\text{A}}\right)^j\left[\pi\delta(\omega-\omega_{nk})-i\mathcal{P}\frac{1}{\omega-\omega_{nk}}\right].
\end{align}
Carrying out the complex-frequency integration separates resonant (pole) and non-resonant contributions:
\begin{align}\nonumber
&C_{nk}^\text{res}=-\frac{i}{64\pi^2\hbar\varepsilon_0z_\text{A}^3}\sum_{j=0}^\infty\frac{(j+2)!}{j!}\int\limits_0^{2\pi}\!\!d\phi \,\bm{d}_{nk}^{(\phi)2}\left(-\frac{ivf_{\phi,\theta}}{2z_\text{A}}\right)^j\\\label{resT}
&\quad\times r_\text{p}^{(j)}(\omega_{nk})\,,\\\nonumber
&C_{nk}^\text{nres}=\frac{i}{128\pi^3\hbar\varepsilon_0z_\text{A}^3}\sum_{j=0}^\infty(j+2)!\int\limits_0^{2\pi}\!\!d\phi\,\bm{d}_{nk}^{(\phi)2}\left(\frac{ivf_{\phi,\theta}}{2z_\text{A}}\right)^j\\
&\quad\times\int\limits_0^\infty\!\!d\xi\frac{(\omega_{nk}+i\xi)^{(j+1)}+(\omega_{nk}-i\xi)^{(j+1)}}{(\omega_{nk}^2+\xi^2)^{(j+1)}}r_\text{p}(i\xi)\,.
\end{align}
This can be rewritten as
\begin{align}\label{resfin}
&C_{nk}^\text{res}
=-\frac{i\,\Theta(\omega_{nk})}{8\pi^2\hbar\varepsilon_0}\!\int\limits_0^{2\pi}\!\!d\phi\!\!\int\limits_0^\infty\!\!d\kappa\kappa^2e^{-2\kappa z_\text{A}}\bm{d}_{nk}^{( \phi)2}r_\text{p}(\omega'_{nk})\,,
\end{align}
and
\begin{align}\label{nresfin}
&C_{nk}^\text{nres}
=\frac{i}{8\pi^3\hbar\varepsilon_0}\int\limits_0^{2\pi}\!\!d\phi\!\!\int\limits_0^\infty\!\!d\kappa\kappa^2e^{-2\kappa z_\text{A}}\bm{d}_{nk}^{( \phi)2}\int\limits_0^\infty\!\!d\xi\frac{\omega'_{nk}r_\text{p}(i\xi)}{\omega_{nk}^{\prime2}+\xi^2}.
\end{align}
The above derivation demonstrates that Eq.~\eqref{CnkMainEq} is valid for either sign of $v_\perp$, as long as the component of velocity away from the interface is not too large. More precisely as long as $v\omega_{nk}\!\ll\!2z_\text{A}$.

\end{document}